\documentclass[prl,aps,twocolumn,showpacs,preprintnumbers,amsmath,amssymb]{revtex4}
\usepackage{dcolumn}
\usepackage{graphicx}
\usepackage{color}

\def\cred{\color{red}}

\begin{document}

%\preprint{APS/123-QED}
\title{The nature of Itinerant Ferromagnetism of SrRuO$_{3}$ : A DFT+DMFT Study}
\author{Minjae Kim and B. I. Min}
\affiliation{Department of Physics, PCTP,
	Pohang University of Science and Technology, Pohang, 790-784, Korea}
\date{\today}

\begin{abstract}
We have investigated the temperature ($T$)-dependent evolution of electronic structures
and magnetic properties of an itinerant ferromagnet SrRuO$_{3}$,
employing the combined scheme
of the density functional theory and the dynamical mean-field theory (DFT+DMFT).
The inclusion of finite dynamical correlation effects beyond the DFT
well describes not only the incoherent hump structure observed
in the photoemission experiment
but also the $T$-dependent magnetic properties in accordance with experiments.
We have shown that the magnetization
of SrRuO$_{3}$ evolves with the Stoner behavior below
the Curie temperature ($T_{c}$),
reflecting the weak itinerant ferromagnetic behavior,
but the local residual magnetic moment
persists even above $T_{c}$,
indicating the local magnetic moment behavior.
We suggest that the ferromagnetism of SrRuO$_{3}$ has dual nature of
both weak and local moment limits,
even though the magnetism of SrRuO$_{3}$ is more itinerant than that of Fe.
\end{abstract}


\maketitle
%==============================================================================

%\section{Introduction}
Electronic correlation effects in the itinerant ferromagnetism is
one of the most essential subjects in condensed matter physics.\cite{Moriya}
In general, there are two limits in describing the itinerant ferromagnetism.
The one is the weak ferromagnetic limit,
in which the magnetic behavior below the Curie temperature ($T_{c}$)
is well described by the Stoner theory.
In this limit, the correlation effect is not so strong,
and the Curie-Weiss susceptibility above $T_{c}$ is
explained by the spin-fluctuation in the reciprocal space.
The other is the local moment limit,
in which the magnetic behavior
is well described by the Heisenberg theory.
The Curie-Weiss susceptibility above $T_{c}$ is explained by
the localized spin in the real space.
For example, ZrZn$_{2}$ is thought to be close to the weak ferromagnetic limit,
while Fe to the local moment limit.
Many itinerant ferromagnets, however,
seem to be in the intermediate
regime between those two limits.
Usually, both longitudinal and transverse spin-fluctuations
have been taken into account to describe itinerant ferromagnets
in the intermediate regime.\cite{Korenman,Uhl,Ruban}

SrRuO$_{3}$ is a well-known itinerant ferromagnet,
which has numerous interesting properties.\cite{Koster}
The fundamental question on SrRuO$_{3}$ is
whether its magnetism belongs to the weak ferromagnetic limit
or to the local moment limit.
Low temperature ($T$) electronic structures and magnetic properties
of SrRuO$_{3}$ are well described by the standard
density functional theory (DFT)-based
band structure calculations.\cite{Mazin,Fujioka,Etz}
However,
there are also several experimental indications of
the sizable electronic correlation effects and
the localized magnetic moment in the real space.
Photoemission (PES) data show that there is an incoherent hump spectrum
at 1 eV below the Fermi level (E$_F$), which is a signature of
the non-negligible electronic correlation strength $U$.\cite{Toyota,Takizawa}
According to magneto-optical and angle-resolved photoemission spectroscopy (ARPES)
experiments,
there remains a finite exchange splitting up to $T_{c}$ ($\sim$160 K),
reflecting the local magnetic moment behavior.\cite{Dodge,Shai}
Indeed, the ratio of the Curie-Wiess moment and
the saturation moment, $P_{c}/P_{s}$, is close to one ($\sim$1.3),
implying the existence of local magnetic moment.\cite{Kanbayasi}
The $T$-derivative of the magnetic part of the resistivity also
shows the local magnetic moment above $T_{c}$.\cite{Klein}
According to recent $T$-dependent optical data,\cite{Jeong}
the magnetic moment below $T_{c}$ evolves with Stoner behavior,
but, above $T_{c}$, there is a finite short-range ordered
localized magnetic moment that drops linearly with increasing $T$.

These experimental indications suggest that,
to describe the itinerant ferromagnetism of SrRuO$_{3}$,
one may have to take into account the dual nature of Ru $4d$ electrons,
itineracy and localization.
The DFT+$U$ method considering the static electronic correlation $U$
was used to describe the incoherent hump feature in PES experiments.\cite{Rondinelli}
The DFT+$U$, however, cannot describe both the coherent and incoherent spectra,
simultaneously.
In view of the dual nature of Ru $4d$ electrons in SrRuO$_{3}$,
the electronic correlation effect should be included dynamically.
Thus, the combination of the DFT and the dynamical mean-field theory (DFT+DMFT) will be
a good choice for the realistic description of the electronic structure
of SrRuO$_{3}$.\cite{Kotliar}
Actually, both the coherent and incoherent structures in the PES of Ru $4d$ bands
were reported to be well described by the DMFT.\cite{Jakobi}
However, the important $T$-dependent evolution of
electronic structures and magnetic properties of SrRuO$_{3}$
has not been explored yet, which is the subject of the present study.

In this Letter, we have shown that the inclusion of the dynamical
correlation effect of Ru 4$d$ electrons in SrRuO$_{3}$ successfully
describes not only its correlated electronic structure but also the $T$-dependent
evolution of electronic structure and magnetism in accordance with experiments.
We have confirmed that, below $T_{c}$, the magnetic moment evolves
with the Stoner behavior, reflecting the itinerant ferromagnetism.
In addition, we have shown that there is a localized magnetic moment,
which remains even above $T_{c}$ and diminishes linearly
upon heating.

We have performed the DFT+DMFT calculations
based on the projection-embedding scheme,
as implemented in Ref.\cite{Haule1}.
The correlated Ru 4$d$ electrons were treated dynamically by the
DMFT local self-energy with t$_{2g}$ and e$_{g}$ orbital projections,
while the rest $s$ and $p$ electrons were treated on the DFT level.
The DFT part calculation was done by using the full-potential
linearized augmented plane wave (FLAPW) band method.\cite{FLAPW,Blaha}
As an impurity solver,
we used the continuous time quantum Monte Carlo method,
and performed the analytical continuation.\cite{Haule1,Haule2,Werner}
For Coulomb interaction parameters, $U$ and $J$,
we have employed $U$=3 eV and $J$=0.6 eV,
as in existing DMFT calculations,\cite{Jakobi}
which describe experimental results
of SrRuO$_{3}$ well.\cite{JSAhn,HDKim,Rondinelli,BKim,Uval}
We have also performed the spin-polarized DFT calculations
to compare the DFT and DFT+DMFT results.
For crystal structure of SrRuO$_{3}$,
we have adopted
experimental structure of orthorhombic \textsl{Pnma} space group.\cite{Jones}
See the Supplemental Material (SM) for computational details
and the crystal structure.\cite{Supp}

\begin{figure}[t]
\includegraphics[width=8cm]{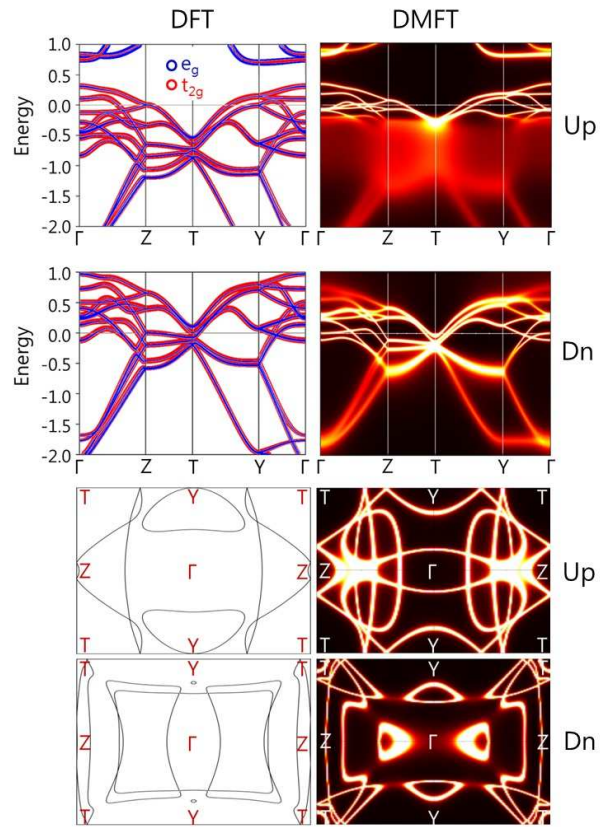}
\caption{(Color online)
(Top) Band structures obtained by the DFT (left)
and by the DFT+DMFT at $T=100$ K (right).
The size of red (blue) dot in the DFT denotes
the amount of t$_{2g}$ (e$_{g}$) component in the wave function.
(Bottom) The DFT FSs (left) and
the DFT+DMFT FSs at $T=100$ K (right).
Up and Dn denote spin up and down, respectively.
See the SM for the first BZ.\cite{Supp}
}
\label{fig1}
\end{figure}

Figure \ref{fig1} presents the band structures obtained by the DFT
and DFT+DMFT schemes at $T$=100 K.
In fact, the DFT+DMFT results represent the spectral functions $A(k,\omega)$.
The seemingly complicated band structures of Ru t$_{2g}$ and e$_g$ arise
from the band folding from the cubic to the orthorhombic Brillouin Zone (BZ).
This feature is consistent with recent report for a similar ruthenate, CaRuO$_{3}$,
which has the same \textsl{Pnma} space group.\cite{Dang1}
There are interesting differences between DFT and DFT+DMFT.
In the DFT+DMFT, one can notice the followings:
(i) the narrowing of the coherent band near E$_F$,
(ii) the emergence of the incoherent feature below E$_F$,
especially for spin up band, and
(iii) the reduction of the exchange splitting.
These features arise from the electronic correlation effects,
and have also been observed in the previous DFT+DMFT calculations
for Fe and Ni.\cite{Lichtenstein}

The flat bands in the vicinity of E$_F$ in Fig. \ref{fig1}
are the Ru t$_{2g}$ bands.
It is seen that, due to the correlation effects,
the DFT+DMFT spin-up (majority spin) bands below $-0.3$ eV become mostly incoherent,
and those above $-0.3$ eV that are coherent become narrowered with respect
to the corresponding DFT bands.
In contrast, the DFT+DMFT spin-down bands show just overall band narrowing
and weak incoherent feature near the top.
The more incoherent feature in the spin-up band than in spin-down band
has already been seen in previous experimental and theoretical studies
on itinerant ferromagnets,
such as Fe, Co, and Ni.\cite{Lichtenstein,Monastra,Grechnev}
Due to almost filled spin-up t$_{2g}$ bands, the electron-hole pair scatterer
is composed of mainly spin-down electrons.
Then the stronger Coulomb scattering between spin up and down electrons
will produce the larger self-energy for the spin-up electrons,
which brings about the stronger incoherent feature in the spin-up band.
Also notable is that, in the DFT+DMFT,
the spin-up and down bands are slightly shifted up and down, respectively,
and so the exchange splitting becomes smaller than that in the DFT.

%%%%%%%%%%%%%%%%%%%%%%%%%%%%%%%%%%%%%%%%%%%%%%%%%%%%%%%%%%%%%%%%
%%%%%%%%%%%%%%%%%%%%%%%%%%%%%%%%%%%%%%%%%%%%%%%%%%%%%%%%%%%%%%%%

In the lower panel of Fig. \ref{fig1},
Fermi surfaces (FSs) obtained by the DFT and DFT+DMFT schemes are provided.
Seemingly different FSs between the DFT and the DFT+DMFT
arise from the smaller exchange splitting in the DFT+DMFT.
In the DFT+DMFT spin-up FSs, due to the shift-up of bands,
electron pockets become smaller, hole pockets become larger,
and new hole pockets emerge, as compared to those in the DFT.
For example, the electron pocket at k=T becomes smaller,
the hole pocket at k=Z becomes larger, and new hole pockets
appear between $\Gamma$ and Z.
Just the opposite situation occurs for the DFT+DMFT spin-down FSs,
due to the shift-down of bands.
For example, the electron pocket between $\Gamma$ and Y becomes larger,
the hole pocket between $\Gamma$ and Z becomes smaller,
and new electron pockets appear at k=T.

%%%%%%%%%%%%%%%%%%%%%%%%%%%%%%%%%%%%%%%%%%%%%%%%%%%%%%%%%%%%%%%%
\begin{figure}[t]
\includegraphics[width=8.5cm]{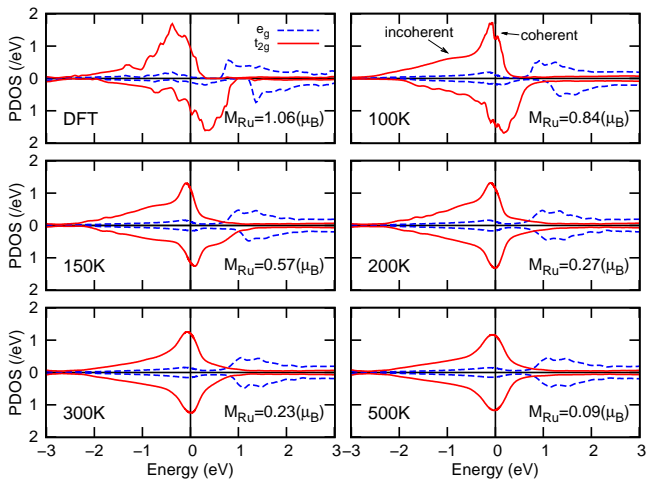}
\caption{(Color online)
Ru $4d$ partial density of states (PDOS)
from the DFT and the DFT+DMFT at different $T$.
Red solid and blue dashed lines represent t$_{2g}$ and e$_{g}$
projected PDOSs, respectively.
Calculated magnetic moment M$_{Ru}$ at each $T$ is provided.
For the DFT+DMFT at $T$=100K,
the coherent and incoherent parts are indicated.
}
\label{fig2}
\end{figure}
%%%%%%%%%%%%%%%%%%%%%%%%%%%%%%%%%%%%%%%%%%%%%%%%%%%%%%%%%%%%%%%%

Comparison of the DFT+DMFT results in Fig. \ref{fig1}
to ARPES data provides
valuable information on the correlation effects.
The spin-up hole pocket FS at k=Z
is much larger in the DFT+DMFT than in the DFT.
The large hole pocket at k=Z is indeed observed in the ARPES experiment.\cite{Shai}
As shown in the SM, the orbital character of this hole pocket at k=Z is $xy$ and $yz$,
which is consistent with the suggestion
from the ARPES experiment.\cite{Shai,Supp,Pbnm}
On the other hand, ARPES shows a kink feature in the
spectral function at 65 meV below E$_F$ along $\Gamma$ and Z.\cite{Shai}
This kink was considered to arise from the electron-boson interaction,
which was proposed as an origin of the small quasi-particle
weight, $Z\sim0.3$.
But, in our DFT+DMFT results, there is no clear kink feature near E$_F$
and the obtained $Z$'s below $T_{c}$ are 0.63 and 0.88 for the spin-up and down bands,
respectively (see Fig.~\ref{fig4}(d)),
which are much larger than the experimental $Z\sim0.3$.\cite{Allen}
Our results of $Z$'s are consistent with previous DFT+DMFT results.\cite{Granas}
This indicates that the kink near E$_F$
and the additional quasi-particle renormalization
might not originate from the local electron-electron interaction, but from the
other electron-boson interaction, such as phonon or $q$-dependent magnon
that requires a scheme beyond the single-site DMFT.

Figure \ref{fig2} presents the partial densities of states (PDOSs)
of t$_{2g}$ and e$_{g}$ orbitals of Ru 4$d$ electrons
in the DFT and DFT+DMFT schemes.
The PDOSs manifest the low-spin state of Ru 4$d$ electrons
with small occupancy of e$_{g}$ orbital.
Note that the exchange splitting in the DFT+DMFT decreases upon heating.
Interestingly, there still remains
a finite exchange splitting near $T_{c}$ ($T\sim$150 K),
as is consistent with magneto-optical\cite{Dodge}
and ARPES experiments.\cite{Shai}

\begin{figure}[t]
\includegraphics[width=8cm]{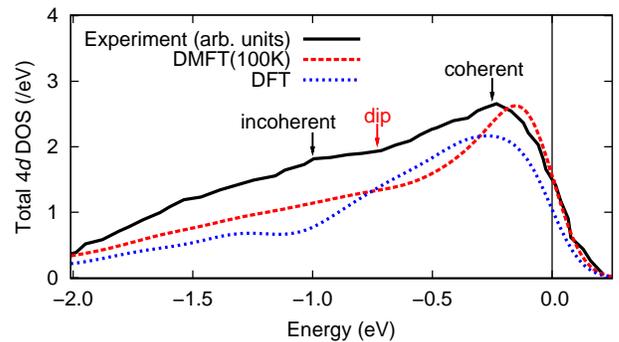}
\caption{(Color online)
The Ru 4$d$ PES spectrum (black solid)\cite{Toyota,Rondinelli}
is compared with the calculated PDOSs in the DFT and the DFT+DMFT (100K).
Blue dotted and red dashed lines
represent the DFT and the DFT+DMFT PDOSs, respectively,
which are broadened with a Gaussian function (0.10 eV FWHM).
The coherent part at $-$0.25 eV, the dip at $-0.73$ eV,
and the incoherent part at $-$1 eV in the PES data are indicated.
}
\label{fig3}
\end{figure}

\begin{figure*}[t]
\includegraphics[width=16cm]{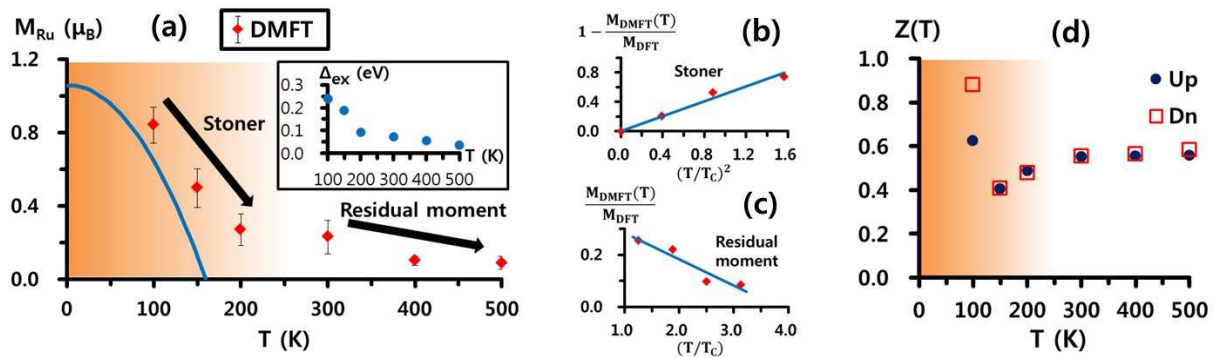}
\caption{(Color online)
(a)
The DFT+DMFT results for the Ru magnetic moment (M$_{Ru}$) in SrRuO$_{3}$
versus $T$.
The blue line represents M$_{Ru}(T)$ with the Stoner behavior
below $T_{c}=160$ K (M$_{Ru}(0)=1.06\mu_{B}$ from the DFT).
We assign the Stoner regime for $T < 200$ K and the residual moment regime
for $T > 200$ K. In the former, M$_{Ru}(T)$ shows the fast drop with $T$,
while, in the latter, it shows the slow drop.
We used averaged M$_{Ru}$ values from the DMFT iterations and the error bar
corresponds to the standard deviation from the averaged value.
Inset presents the $T$-dependent exchange splitting ($\Delta_{ex}$)
between spin up and down band,
which shows the similar behavior to M$_{Ru}(T)$.
(b) The reduction of scaled magnetic moment of
the DFT+DMFT ($1-\frac{M_{DMFT}}{M_{DFT}}$) versus
the square of scaled temperature ($T/T_{c}$)$^{2}$ in the Stoner regime.
The blue linear line is from the least square fitting of data, including the origin.
(c) The scaled magnetic moment of the DFT+DMFT ($\frac{M_{DMFT}}{M_{DFT}}$)
versus the scaled temperature ($T/T_{c}$) in the residual moment regime.
The blue linear line is from the least square fitting.
(d)
The DFT+DMFT results for the quasi-particle weight ($Z$) of t$_{2g}$ orbital versus $T$.
Blue filled circles and red empty squares represent
$Z$'s for spin-up and down electrons, respectively.
}
\label{fig4}
\end{figure*}

To examine the signature of electronic correlation effects in SrRuO$_{3}$,
Ru 4$d$ PES spectrum\cite{Toyota,Rondinelli} is compared with near-E$_F$ PDOSs of
the DFT and the DFT+DMFT ($T$=100 K) in Fig. \ref{fig3}.
Differently from the DFT PDOS,
the DFT+DMFT PDOS shows both the coherent and the incoherent part
and the shallow dip structure in-between,
as is consistent with the experimental PES data.
Interestingly, the peak position of incoherent PDOS, 1 eV below E$_F$,
is consistent with that in the previous DFT+$U$
($U$=1 eV) result.\cite{Toyota,Takizawa,Rondinelli}
The consistency between the DFT+$U$ and DFT+DMFT results
justifies the use of a bit larger $U$=3 eV in the DMFT.\cite{Uval}

%%%%%%%%%%%%%%%%%%%%%%%%%%%%%%%%%%%%%%%%%%%%%%%%%%%%%%%%%%%%%%%%
%%%%%%%%%%%%%%%%%%%%%%%%%%%%%%%%%%%%%%%%%%%%%%%%%%%%%%%%%%%%%%%%

%%%%%%%%%%%%%%%%%%%%%%%%%%%%%%%%%%%%%%%%%%%%%%%%%%%%%%%%%%%%%%%%
%%%%%%%%%%%%%%%%%%%%%%%%%%%%%%%%%%%%%%%%%%%%%%%%%%%%%%%%%%%%%%%%

The finite dynamical Coulomb correlation in the DFT+DMFT also reproduces
the correct $T$-dependent magnetic properties of SrRuO$_{3}$.
Figure \ref{fig4}(a) presents the DFT+DMFT results of the
Ru magnetic moment (M$_{Ru}$) in SrRuO$_{3}$ versus $T$,
which shows that there are two regimes
for the evolution of M$_{Ru}$ with increasing $T$.
Below $T=200$ K, M$_{Ru}$ drops rather rapidly upon heating.
On the other hand, above $T=200$ K,
M$_{Ru}$ diminishes rather slowly upon heating.
Noteworthy is that there
remains a significant local magnetic moment near $T_{c}$
(0.27$\mu_{B}$ at $T$=150 K),
which seems to persist even above $T_{c}$.
According to the Stoner theory,
the local magnetic moment at $T$=0
is supposed to be close to the DFT value (1.06 $\mu_{B}$),
and it drops to zero at $T_{c}$ with the $T^{2}$ scaling behavior,
as represented by the blue line Fig. \ref{fig4}(a).
Thus the M$_{Ru}-T$ curve in Fig. \ref{fig4}(a) indicates that
the Stoner theory of the itinerant magnetism is not
appropriate in describing the $T$-dependent evolution of
the local magnetic moment in SrRuO$_{3}$ above $T_{c}$.
It is seen that the $T$-dependent behavior is different
below and above $T=200$ K.
We designate the $T$ regimes below and above $T=200$ K as
the Stoner and residual moment regimes, respectively.

Figure \ref{fig4}(b) and (c) present $T$-dependent behaviors
of magnetic moment in the Stoner and the residual moment regime, respectively.
In the Stoner regime below $T_{c}$, Fig. \ref{fig4}(b) clearly
shows that the magnetic moment diminishes with the $T^{2}$ scaling,
which is in good agreement with the $T$-dependent spectrum evolution
in optical experiment.\cite{Jeong}
On the other hand, in the residual moment regime in Fig. \ref{fig4}(c),
the finite local magnetic moment even above $T_{c}$ is seen to diminish
linearly with $T$,
which is again consistent with the optical spectrum experiment.\cite{Jeong}
The $T$-dependent exchange splitting ($\Delta_{ex}$) of the t$_{2g}$ band
in the inset of Fig.\ref{fig4}(a) also follows the $T^{2}$ and
$T$ scaling behaviors below and above $T_{c}$, respectively.
The disappearance of the net macroscopic magnetic moment above $T_{c}$
would be driven by the long wave length transversal spin wave
excitations of the persistent local residual moment.\cite{Jeong,Lichtenstein}

The linearly diminishing behavior of the local magnetic moment above $T_{c}$ in
SrRuO$_{3}$ is reminiscent of the theoretically suggested
behavior in Fe, which was derived by using
the Monte Carlo simulation with inclusion of both
longitudinal and transverse spin-fluctuations.\cite{Ruban}
This kind of simulation is in line with the intermediate approach between
the weak ferromagnetic limit and the local moment limit.
But, in the case of SrRuO$_{3}$, our result shows that
the persistent local moment above $T_{c}$ is smaller and
diminishes more rapidly with increasing $T$ than in Fe.
This difference indicates that
the magnetic properties of SrRuO$_{3}$ are still more itinerant than those of Fe.

We have confirmed that the ferromagnetic ordering restores
the coherent metallic nature in SrRuO$_{3}$.\cite{Georges,Dang2}
Figure \ref{fig4}(d) presents the $T$-dependent quasi-particle weight, $Z$,
of t$_{2g}$ orbital.
$Z$ is estimated from the analytically continued self-energy of t$_{2g}$
orbital in the DFT+DMFT
($Z=(1-\frac{\partial}{\partial\omega}Re\Sigma(\omega))^{-1}\mid_{\omega=0}$).
Above $T_{c}$ ($\sim$150K), the increasing spin-flip scattering upon cooling
reduces the coherence of t$_{2g}$ orbital, as shown in the
diminishing $Z$ in Fig. \ref{fig4}(d).
On the other hand, below $T_{c}$,
the spin-flip scattering is suppressed with ferromagnetic ordering,
as shown in the enhancement of $Z$ upon cooling.
Therefore, the crossover occurs near $T_{c}$
from the spin-flip induced incoherent regime
to the ferromagnetic ordering induced coherent regime.

In conclusion,
we have shown that the DFT+DMFT calculations
describes correctly the correlated electronic structure
and the $T$-dependent evolutions of local electronic structure and
magnetic moment in SrRuO$_{3}$.
The obtained Stoner behavior below $T_{c}$ and the persistent
local magnetic moment above $T_{c}$ are consistent with experiments.
The $T$-dependent local magnetic moment evolution in SrRuO$_{3}$
is similar to the case in Fe.
More rapid drop of local magnetic moment in SrRuO$_{3}$  above $T_{c}$, however,
indicates that Ru $4d$ electrons in SrRuO$_{3}$ are more itinerant
than Fe $3d$ electrons.
The present studies of the $T$-dependent electronic structures
and magnetic properties based on the DFT+DMFT method
will provide an important tool to understand fundamental properties
in various itinerant ferromagnets.

{\cred
%After completion of the manuscript, we became aware of a related work by
%Dang \textit{et al.,}\cite{Dang2}.
%The spin-dependent correlation effect and
%the restoration of coherence with onset of ferromagnetic order in SrRuO$_{3}$
%are consistent with our results.
}

%\begin{acknowledgments}
Helpful discussions with C.-J. Kang, Kyoo Kim, and Beom Hyun Kim are greatly
appreciated.
This work was supported by the POSTECH BSRI grant
and the KISTI supercomputing center (No. KSC-2013-C3-010).
%\end{acknowledgments}


\begin{thebibliography}{99}
\bibitem{Uval} The $U$ value of 3 eV is apparently somewhat larger than
that used in the DFT+$U$ method ($\sim$1 eV).\cite{Rondinelli,BKim}
The reason of using larger $U$ in the DMFT
is that the given bare $U$ is to be reduced by the screening channel
in the DMFT impurity solver.
\bibitem{Pbnm} The $y$ axis in the \textsl{Pnma} space group is
the $z$ axis in the \textsl{Pbnm} space group
which is the [001] axis of the SrRuO$_{3}$ film on the substrate.\cite{Shai}
\end{thebibliography}
\end{document}